\documentclass{emulateapj}

\usepackage{psfig}
\usepackage{natbib}
\newcommand{\oneprime}{\mbox{$^{\prime}$}}%

\shorttitle{Ly$\alpha$ Pumped H$_{2}$ Emission in PNe}
\shortauthors{Lupu et al.}

\begin{document}


\title{Discovery of Ly$\alpha$ Pumped Molecular Hydrogen Emission
in the Planetary Nebulae NGC 6853 and NGC 3132}


\author{R. E. Lupu, K. France,  and S. R. McCandliss}
\affil{Department of Physics and Astronomy, Johns Hopkins University,
    Baltimore, MD 21218}





\begin{abstract}

We report the first observation of Ly$\alpha$ pumped molecular hydrogen emission lines in planetary nebulae. 
The H$_{2}$ emission observed in the ultraviolet spectra of NGC 6853 and NGC 3132 can be explained by 
excitation of vibrationally hot H$_{2}$ by Ly$\alpha$ photons. Constraints are placed on the nebular Ly$\alpha$ emission profile, as well as the molecular hydrogen temperature, column density and turbulent motion. These parameters are similar for the two nebulae, pointing to similar physical conditions in these objects. The ro-vibrational cascade following Ly$\alpha$ pumping is predicted to have low surface brightness signatures in the visible and near infrared. 

\end{abstract}


\keywords{ISM:molecules~---~planetary nebulae:individual~(NGC 6853, NGC 3132)~---~
	~ultraviolet:ISM}

\section{INTRODUCTION}
\hspace{0.25in}

Molecular hydrogen emission is often seen in the infrared (IR) observations of photodissociation 
regions within planetary nebulae (PNe) \citep{zuckerman88,kastner96}. Continuum pumped ultraviolet (UV) fluorescence, as observed in reflection nebulae and star forming regions \citep{witt89,france05b}, has not been detected in PNe. The observed IR quadrupole emission originates in elevated ro-vibrational levels of the ground electronic state $X$$^{1}\Sigma^{+}_{g}$. The distribution of molecules in these levels is characteristic of either shock heating \citep{zuckerman88} or radiative cascade from excited electronic states \citep{hayashi85,gatley87}. UV fluorescence can result in the dissociation of the molecule 10 -- 15$\%$ of the time \citep{stecher67}, or a non-thermal population in the excited ro-vibrational levels of $X$$^{1}\Sigma^{+}_{g}$. Shocks, with specific temperatures of $\sim$~2000~K, can thermally populate higher ro-vibrational states, resulting in IR transitions without the far-UV cascade \citep{takami00}. Detection of the UV fluorescence in conjunction with models of the infrared to ultraviolet scaling of H$_{2}$ emission can constrain the contribution of each mechanism to the IR flux. The line pumped H$_{2}$ UV fluorescence described in this paper is consistent with measured IR line ratios and gas temperatures that favor the shock excitation scenario \citep{zuckerman88,storey84}. As shown in \S~4.1, this type of UV emission, unlike continuum pumped fluorescence, has little effect on the observed IR spectrum. Weak specific features are expected in the optical.

At typical diffuse ISM temperatures below 100~K \citep{rachford02}, only H$_{2}$ molecules in the v\arcsec~=~0 level of $X$$^{1}\Sigma^{+}_{g}$ make a significant contribution to the UV fluorescence. These molecules are excited by 912~$\lesssim$~$\lambda$~$\lesssim$~1110~\AA\ radiation into vibrational and rotational levels of higher electronic states (mainly $B$$^{1}\Sigma^{+}_{u}$ and $C$$^{1}\Pi_{u}$, Black \& van Dishoeck 1987). As the temperature increases, higher ro-vibrational levels of $X$$^{1}\Sigma^{+}_{g}$ become populated, and the threshold for fluorescent pumping moves to progressively larger wavelengths. Due to the long lifetime of the $X$$^{1}\Sigma^{+}_{g}$ levels, the molecules can be further pumped to $B$$^{1}\Sigma^{+}_{u}$ or $C$$^{1}\Pi_{u}$ by a sufficiently strong radiation field before they have time to radiate. This mechanism is predicted to be important for nebular gas densities in the range from 10$^4~\le$~n~$\le~$10$^6$~cm$^{-3}$ and a radiation field (in units of the interstellar average) of 10$^3~\le$~G$_0~\le$~10$^4$ \citep{sternberg89}.   
 
We have found that in both NGC~6853 (M27, The Dumbbell Nebula) and NGC~3132 H$_{2}$ UV fluorescence is dominated by lines pumped by nebular Ly$\alpha$ from the excited v\arcsec~=~2 level of $X$$^{1}\Sigma^{+}_{g}$ to the $B$$^{1}\Sigma^{+}_{u}$ states. Molecular hydrogen resonance fluorescence with Ly$\alpha$ has been previously recognized as an important contributor to the UV flux in collisional environments of photodissociation regions. 
This effect was first pointed out by \citet{shull78}, and discussed further by \citet{black87}.  
Ly$\alpha$ pumping was predicted to be significant mainly for T Tauri stars and Herbig-Haro 
objects, as has been subsequently observed \citep{brown81,schwartz83,raymond97,herczeg04}. Other environments in which Ly$\alpha$ pumping was detected include solar system objects \citep{wolven97,jordan77} and accreting systems \citep{gizis05,wood02}. In this paper we 
present for the first time evidence for resonant excitation by Ly$\alpha$ of the $B-X$~(1~--~2)~P(5) 1216.07~\AA\ and $B-X$~(1~--~2)~R(6) 1215.73~\AA\ molecular hydrogen lines in planetray nebulae. The observed resonances 
and line ratios are a valuable diagnostic tool for molecular gas temperature, Ly$\alpha$ line shape, and
Doppler shift relative to the H$_{2}$ absorber.

The first object, NGC~6853, is a 12,700 year old planetray nebula \citep{odell02} at a distance of 
417~$\pm$~50~pc \citep{benedict03}. Its central star has a temperature 108,600~$\pm$~6800~
K \citep{benedict03}. The distance to NGC~3132 is not well known, varying from 1.63~kpc 
\citep{torres77} to 0.51~kpc \citep{pottasch96}, while the temperature of the ionizing star is 
estimated at 110,000 K \citep{pottasch96}. The extinction, as measured by $E(B-V)$ in this case, is of the order of 0.1 for the 
central star in both cases \citep{benedict03,pottasch96}. However, \citet{mcc06} argue that in the 
case of NGC~6853 the extinction is negligible. In both cases, the detected Ly$\alpha$ pumped H$_{2}$ fluorescence requires a temperature of $\sim$~2000~K and column densities of $\sim$~10$^{18}$~cm$^{-2}$, as well as a Ly$\alpha$ profile of $\sim$~0.4~\AA\ width, with a deep self-reversal. These parameters satisfy both 
short and long wavelength constraints, pointing to similar conditions of the molecular gas in the two 
nebulae.   

A description of the observations and data is found in Section 2. Data analysis follows in Section 
3, and a discussion of the results in Section 4.
 
 \begin{deluxetable*}{cccccc}
\tabletypesize{\small}
\tablecaption{FUSE Observations of NGC 6853. \label{table1}}
\tablewidth{0pt}
\tablehead{
\colhead{Instrument} & \colhead{Program}   & \colhead{Date}   &
\colhead{RA} & \colhead{Dec} & \colhead{Exp. Time (s)}\\ 
}
\startdata
FUSE - LWRS             &     E12001   &    2004-05-26  & 19 59 44.96    &  +22 44 50.1  & 4176.0 
\\
FUSE - LWRS            &    E12002   &    2004-05-26 &  19 59 38.93  &    +22 44 04.0  & 3071.0  \\
FUSE - LWRS             &    E12003   &   2004-05-26  & 19 59 32.35     &   +22 42 35.1  & 2778.0  
\\
FUSE - LWRS              &    E12004   &    2004-05-26 &  19 59 27.27   &   +22 42 10.3  & 3470.0  
\\
\enddata
\end{deluxetable*}

\section{OBSERVATIONS}

\begin{figure}
\begin{center}
\rotatebox{0}{
\plotone{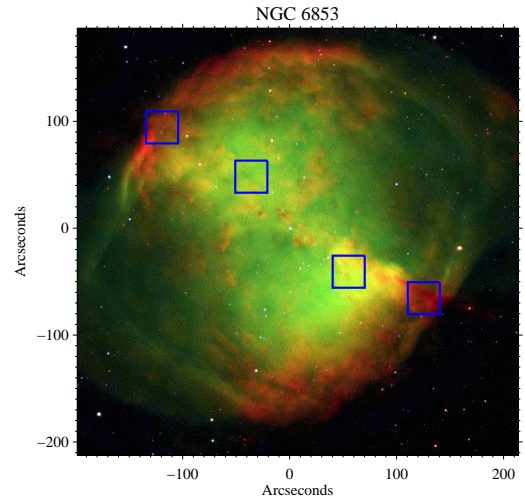} }
\caption{~~Optical image of NGC~6853 obtained during the commissioning of the FORS1
instrument on the VLT 8.2-m.  This three color image was made from
a composite of B, [\ion{O}{3}], and H$\alpha$ filters.  $FUSE$ LWRS
aperture overlays are imposed showing the location of the four
nebular pointings 1 to 4 from left to right.  The optical image was obtained from the
European Southern Observataory.\label{fusem27} }
\end{center}
\end{figure}

Four nebular  
observations of NGC 6853 were made by $FUSE$ on 2004 May 26 using the low-resolution (LWRS) 
aperture 
(30\arcsec$\times$30\arcsec; Figure~\ref{fusem27}).  
Spectra were obtained in the 905~--~1187~\AA\ bandpass at a filled
aperture resolution of $\sim$~0.33~\AA.  A description of the 
$FUSE$ satellite can be found in ~\citet{moos00} and on-orbit
performance characteristics are described by ~\citet{sahnow00}. 
The data were acquired as part of the E120 guest observing
program. The pointings extend along the bright bar
structure of the nebula, from the northeast to the southwest (Figure~\ref{fusem27}).
Data for all four pointings were obtained in ``time-tagged'' (TTAG) mode, 
and processed using the CALFUSE pipeline, version 2.4.2. The average 
exposure time was 3374 seconds. The log of the observations of NGC 6853 is presented in 
Table~\ref{table1}.

We obtained optical spectra of NGC~6853 on 1999 June 9 using the Double
Imaging Spectrograph (DIS) on the Apache Point 3.5m telescope. The observations were taken with the low-resolution gratings, centered at 8000~\AA\ and 4224~\AA\, with a dispersion of 7.0~\AA~pixel$^{-1}$ in the red side and 6.1~\AA~pixel$^{-1}$ in the blue side, and a resolution of roughly 2~pixels. The 300\arcsec$\times$0\farcs9 slit was centered at (19$^{\mathrm h}$~59$^{\mathrm m}$~36$^{\mathrm s}$.20, 22\arcdeg~43\arcmin~01\arcsec.00), and oriented along the bright bar. Three slit positions (about 2\farcs7 total width) were averaged during the 900~s integration. The observed mean H$\alpha$ brightness along the slit was about 4.68~$\times$~10$^{-3}$~ergs~cm$^{-2}$~s$^{-1}$~sr$^{-1}$, or 19416 R (conversion factor 632 $\lambda_{H\alpha}$(\AA)~).

NGC 3132 was observed by HUT aboard $Astro-2$ for 946 seconds
during orbital night on 1995 March 14.  This spectrum, 
downloaded from the Multi-Mission Archive at the 
Space Telescope Science Institute (MAST),
was acquired through the 10\arcsec$\times$56\arcsec~slit, enclosing a large region of the 
28\arcsec\ nebula, at an offset from the
central stars. The exact slit coordinates are uncertain. The resolution of the HUT spectrum is
roughly 3~\AA.  A description of the HUT 
instrument and data reduction can be found in~\citet{kruk95}.
The signature of H$_{2}$ emission is present in the 
$FUSE$ spectrum of NGC 3132 (D1200401), but at a considerably lower 
signal-to-noise ratio. We chose to analyze the HUT spectrum because it allowes us to extend our study to longer wavelengths.

\section{DATA ANALYSIS}

\subsection{NGC 6853}

\begin{figure*}
\begin{center}
\epsscale{0.5}
\rotatebox{90}{
\plotone{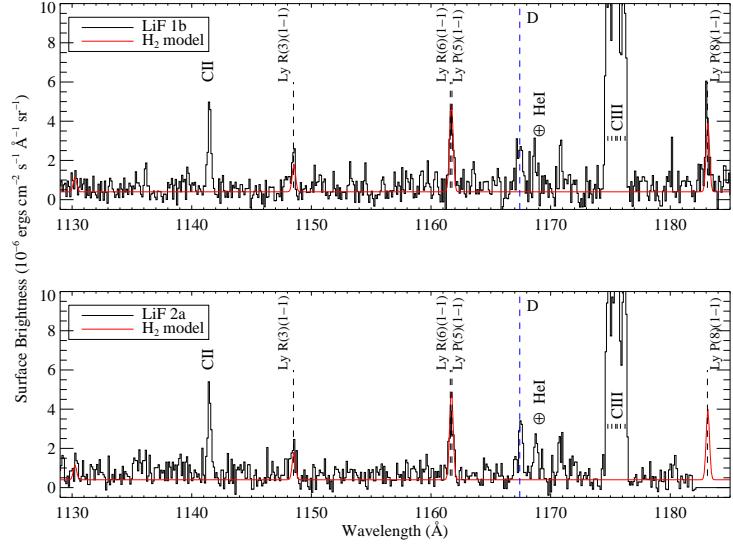} }
\caption{~~ $FUSE$ spectra of NGC~6853 at the third pointing. Both the LiF~1b (upper) and LiF~2a 
(lower) channels are shown to cover the whole 1115~--~1187~\AA\ interval. The H$_{2}$ emission is 
prominent together with \ion{C}{2}, \ion{C}{3} and geocoronal \ion{He}{1}. The Ly$\alpha$ pumped 
H$_{2}$ fluorescence model is shown in red. \label{fusespec}  }
\end{center}
\end{figure*}

\begin{figure}
\begin{center}
\epsscale{0.9}
\rotatebox{90}{
\plotone{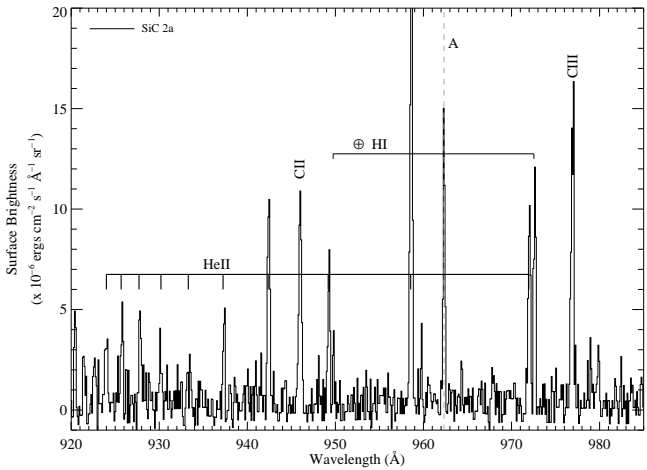} }
\caption{~~ The short wavelength FUSE SiC~2a spectrum of NGC~6853 at the third pointing. Strongest 
features are nebular \ion{He}{2}, \ion{C}{2}, \ion{C}{3} and geocoronal \ion{H}{1}.  
\label{heiip3}  }
\end{center}
\end{figure}

The $FUSE$ spectra of NGC~6853~--~Position 3 are presented in Figure~\ref{fusespec} and 
Figure~\ref{heiip3}. The emission from the nebula is dominated by \ion{He}{2}, \ion{C}{2}, \ion{C}{3} and molecular hydrogen lines. The strongest lines 
observed in all four pointings are summarized in Tables 2-5. The lines denoted by letters $A-D$ are 
unidentified. Feature $A$ is possibly nebular \ion{O}{3} (962.425~\AA) at a slight velocity offset. The strengths of the lines in the spectrum changes with the pointing. The highly ionized 
species dominate in the central regions and decline towards the outer shell. Molecular hydrogen 
lines are present in all four pointings, and are strongest in the third pointing, which coincides with the brightest optical feature.

\begin{deluxetable}{cccc}
\tabletypesize{\small}
\tablecaption{NGC 6853 Position 1. \label{table2}}
\tablewidth{0pt}
\tablehead{
\colhead{Line ID} & \colhead{$\lambda_{obs}$ }      &
\colhead{FWHM} & \colhead{Brightness }\\ 
	&	(\AA)		&	(\AA)	&
	(10$^{-6}$ ergs cm$^{-2}$ s$^{-1}$ sr$^{-1}$)	 
}
\startdata
C II                     &  946.01 &      0.56 &    1.83 $\pm$    0.29\\
He II                    &  958.52 &      0.53 &    0.35 $\pm$    0.28\\
N III + He II            &  991.52 &      0.21 &    1.91 $\pm$    0.39\\
C II                     & 1036.88 &      0.33 &    1.37 $\pm$    0.22\\
N II + He II             & 1085.50 &      0.34 &   25.75 $\pm$    1.03\\
C II                     & 1141.53 &      0.26 &    0.36 $\pm$    0.13\\
H$_{2}$(1--1)R(3)        & 1148.59 &      0.43 &    0.40 $\pm$    0.14\\
H$_{2}$(1--1)P(5) + R(6) & 1161.76 &      0.32 &    0.71 $\pm$    0.16\\
C III                    & 1175.53 &      0.48 &    1.97 $\pm$    0.23\\
H$_{2}$(1--1)P(8)        & 1183.06 &      0.21 &    0.92 $\pm$    0.32\\
\enddata
\tablecomments{The \ion{C}{3} line at 977.03~\AA\ is not listed due to contamination from scattered solar light in the SiC channels.} 
\end{deluxetable}

\begin{deluxetable}{cccc}
\tabletypesize{\small}
\tablecaption{NGC 6853 Position 2. \label{table3}}
\tablewidth{0pt}
\tablehead{
\colhead{Line ID} & \colhead{$\lambda_{obs}$ }      &
\colhead{FWHM} & \colhead{Brightness }\\ 
	&	(\AA)		&	(\AA)	&
	(10$^{-6}$ ergs cm$^{-2}$ s$^{-1}$ sr$^{-1}$)	 
}
\startdata

He II                    &  933.35 &      0.26 &    1.57 $\pm$    0.43\\
He II                    &  937.24 &      0.21 &    0.78 $\pm$    0.37\\
He II                    &  942.43 &      0.29 &    6.15 $\pm$    0.50\\
C II                     &  946.02 &      0.46 &    5.16 $\pm$    0.48\\
He II                    &  949.24 &      0.43 &    2.83 $\pm$    0.38\\
He II                    &  958.59 &      0.29 &   12.37 $\pm$    0.61\\
Un-ID (A)                &  962.31 &      0.27 &    4.24 $\pm$    0.48\\
N III + He II            &  992.30 &      0.31 &   48.83 $\pm$    1.28\\
S III                    & 1015.48 &      0.89 &    1.65 $\pm$    1.08\\
S III                    & 1021.32 &      0.68 &    0.72 $\pm$    0.25\\
C II                     & 1036.90 &      0.42 &    3.59 $\pm$    0.34\\
Un-ID (B)                & 1062.62 &      0.44 &    0.31 $\pm$    0.28\\
Un-ID (C)                & 1065.84 &      0.47 &    0.98 $\pm$    0.27\\
N II + He II             & 1084.79 &      0.32 &  130.94 $\pm$    2.46\\
C II                     & 1141.51 &      0.37 &    1.26 $\pm$    0.23\\
H$_{2}$(1--1)R(3)        & 1148.63 &      0.23 &    0.62 $\pm$    0.23\\
H$_{2}$(1--1)P(5) + R(6) & 1161.75 &      0.40 &    1.62 $\pm$    0.23\\
Un-ID (D)                & 1167.30 &      0.51 &    1.24 $\pm$    0.27\\
C III                    & 1175.53 &      1.02 &   45.55 $\pm$    0.78\\
H$_{2}$(1--1)P(8)        & 1183.26 &      0.35 &    2.01 $\pm$    0.51\\
\enddata
\end{deluxetable}

\begin{deluxetable}{cccc}
\tabletypesize{\small}
\tablecaption{NGC 6853 Position 3. \label{table4}}
\tablewidth{0pt}
\tablehead{
\colhead{Line ID} & \colhead{$\lambda_{obs}$ }   &
\colhead{FWHM} & \colhead{Brightness }\\ 
	&	(\AA)		&	(\AA)	&
	(10$^{-6}$ ergs cm$^{-2}$ s$^{-1}$ sr$^{-1}$)	 
}
\startdata
He II                    &  933.40 &      0.54 &    1.21 $\pm$    0.36\\
He II                    &  937.30 &      0.26 &    0.91 $\pm$    0.34\\
He II                    &  942.46 &      0.30 &    3.63 $\pm$    0.44\\
C II                     &  946.03 &      0.39 &    4.75 $\pm$    0.44\\
He II                    &  949.30 &      0.22 &    2.06 $\pm$    0.33\\
He II                    &  958.63 &      0.29 &    8.86 $\pm$    0.54\\
Un-ID (A)                &  962.35 &      0.27 &    3.69 $\pm$    0.45\\
N III + He II            &  992.34 &      0.32 &   30.27 $\pm$    1.02\\
S III                    & 1015.42 &      0.81 &    1.20 $\pm$    0.32\\
S III                    & 1021.16 &      0.52 &    0.57 $\pm$    0.20\\
C II                     & 1036.81 &      0.33 &    2.37 $\pm$    0.29\\
Un-ID (B)                & 1062.34 &      0.31 &    0.20 $\pm$    0.25\\
Un-ID (C)                & 1065.79 &      0.30 &    1.40 $\pm$    0.26\\
N II + He II             & 1084.82 &      0.34 &   95.18 $\pm$    2.18\\
C II                     & 1141.48 &      0.31 &    1.50 $\pm$    0.22\\
H$_{2}$(1--1)R(3)        & 1148.48 &      0.42 &    0.68 $\pm$    0.23\\
H$_{2}$(1--1)P(5) + R(6) & 1161.70 &      0.42 &    1.33 $\pm$    0.22\\
Un-ID (D)                & 1167.44 &      0.50 &    1.65 $\pm$    0.26\\
C III                    & 1175.48 &      1.07 &   28.33 $\pm$    0.65\\
H$_{2}$(1--1)P(8)        & 1183.05 &      0.26 &    3.17 $\pm$    0.49\\
\enddata
\end{deluxetable}

\begin{deluxetable}{cccc}
\tabletypesize{\small}
\tablecaption{NGC 6853 Position 4. \label{table5}}
\tablewidth{0pt}
\tablehead{
\colhead{Line ID} & \colhead{$\lambda_{obs}$ }   &
\colhead{FWHM} & \colhead{Brightness }\\ 
	&	(\AA)		&	(\AA)	&
	(10$^{-6}$ ergs cm$^{-2}$ s$^{-1}$ sr$^{-1}$)	 
}
\startdata
C II                     &  946.05 &      0.38 &    1.22 $\pm$    0.30\\
He II                    &  958.59 &      0.24 &    0.34 $\pm$    0.32\\
N III + He II            &  991.63 &      0.32 &    1.56 $\pm$    0.42\\
C II                     & 1037.75 &      0.42 &    0.56 $\pm$    0.22\\
N II + He II             & 1085.54 &      0.34 &   18.76 $\pm$    0.97\\
C II                     & 1141.52 &      0.25 &    0.21 $\pm$    0.14\\
H$_{2}$(1--1)R(3)        & 1148.50 &      0.27 &    0.12 $\pm$    0.14\\
H$_{2}$(1--1)P(5) + R(6) & 1161.73 &      0.21 &    0.66 $\pm$    0.17\\
C III                    & 1175.56 &      0.29 &    1.29 $\pm$    0.23\\
H$_{2}$(1--1)P(8)        & 1183.18 &      0.31 &    1.49 $\pm$    0.34\\
\enddata
\end{deluxetable}

We explored the possibility that line-pumped H$_{2}$ fluorescence is responsible for some of the 
unidentified lines we observe, computing models with a variety of nebular emission lines as the 
excitation source. Atomic emission lines of \ion{H}{1} Ly-$\beta$, Ly-$\delta$, and Ly-$\gamma$; 
\ion{He}{2}~$\lambda$1215, $\lambda$1085, $\lambda$1025, $\lambda$992, and $\lambda$959; \ion{C}{2} 
$\lambda\lambda$1334/35 and $\lambda\lambda$1036/37; the \ion{C}{3} $\lambda$1175 
multiplet and \ion{C}{3} $\lambda$977; \ion{N}{1} $\lambda$1200; the \ion{N}{2} $\lambda$1085 multiplet; the \ion{N}{3} 
$\lambda$990 multiplet; and the \ion{O}{6} resonance doublet $\lambda\lambda$1032/38 were all 
considered as possible H$_{2}$ excitation mechanisms. None of the strongest predicted fluorescent 
emission lines are observed. 

\begin{figure}
\begin{center}
\epsscale{0.9}
\rotatebox{90}{
\plotone{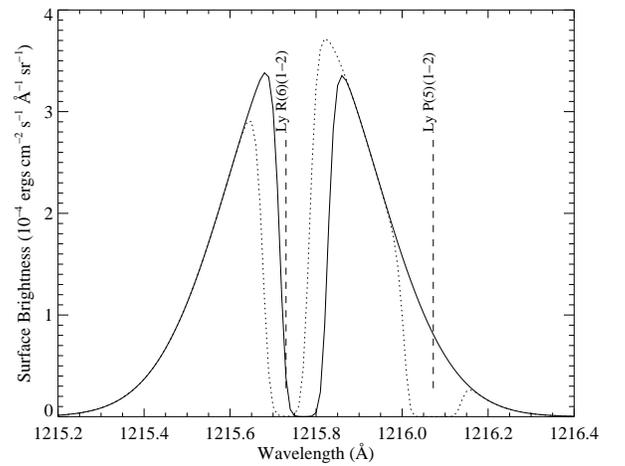} }
\caption{~~ Exciting Ly$\alpha$ profile absorbed by a \ion{H}{1} column density of 
1~$\times$~10$^{14}$~cm$^{-2}$ used for the NGC~6853 model. Velocity offsets with respect to the 
molecular hydrogen reference system are 25~km~s$^{-1}$ for both emission and \ion{H}{1} absorption. The 
H$_{2}$ absorbing transitions are indicated by dashed lines. The dotted line corresponds to an H$_{2}$ absorption profile for a column density of 6~$\times$~10$^{18}$~cm$^{-2}$. \label{lya}  }
\end{center}
\end{figure}

The strongest H$_{2}$ lines observed are excited by the $B-X$ (1~--~2)~P(5) transition at 
1216.07~\AA\, in the red wing of Ly$\alpha$. Their wavelengths and branching ratios are listed in Table~\ref{table_branch}. These transitions are sensitive to the Ly$\alpha$ line 
width and Doppler shift. Although the closest resonance is $B-X$ (1~--~2)~R(6) at 1215.73~\AA, 
Lyman-system fluorescence from v\oneprime~=~1, J\oneprime~=~7 is far weaker than expected. This can be achieved with a deep self-reversal of the Ly$\alpha$ profile combined with a blueshifting of the molecular hydrogen lines with respect to the systemic Ly$\alpha$ emission. The bueshift is necessary to restore to some extent the intensities of the lines pumped by the (1~--~2) R(6) resonance, which otherwise would fall in the center of the absorption trough (Figure~\ref{lya}). 

NGC~6853, as well as NGC~3132, was modeled using a fluorescent H$_{2}$ emission code similar to the one described in 
\citet{france05a}. The exciting radiation field was assumed to be a Gaussian Ly$\alpha$ emission line, with a FWHM of~0.4~\AA. In order to mimic self-reversal, absorption by neutral hydrogen with a column density of 1~$\times$~10$^{14}$~cm$^{-2}$ was used to modify the exciting Ly$\alpha$ profile (Figure~\ref{lya}). As the nebular Ly$\alpha$ line cannot be measured directly due to the opacity of 
the ISM and contamination from geocoroanal Ly$\alpha$, a total brightness was chosen as 2/3 of the H$\alpha$ brightness of 19416.0~R measured using the DIS at APO (see \S~2). According to the selection rules, 2/3 of the time the H$\alpha$ transitions will result in an atom in the 2$p$ state, which can then radiate a Ly$\alpha$ photon during the 2$p$~--~1$s$ transition. The other 1/3 of the time, the hydrogen atom will be left in the 2$s$ state, unable to emit a Ly$\alpha$ photon since a direct transition from this state to 1$s$ is forbidden in the dipole approximation \citep{spitzer78}. The total Ly$\alpha$ brightness could be higher than derived from H$\alpha$ measurements, due to H$_{2}$ dissociation mechanisms that produce \ion{H}{1} in the 2$p$ state \citep{glass86}. The continuum flux from the central star of NGC 6853 has not been included because continuum pumped emission lines are not detected in the nebular spectra. The likely explanation for the absence of continuum pumped fluorescence resides in the dominance of Ly$\alpha$
photons over the 912~--~1110~\AA\ stellar continuum as well as the small covering factor involved in the absorption of continuum radiation, as will be discussed in \S~4.

\begin{deluxetable}{ccc}
\tabletypesize{\small}
\tablecaption{Lyman system transitions pumped by Ly$\alpha$ out of the (v\arcsec,J\arcsec) levels (2,5) and (2,6). \label{table_branch}}
\tablewidth{0pt}
\tablehead{
\colhead{Line ID} &  {Wavelength (\AA)}  &  {Branching Ratio ($\times$~10$^{4}$)}}
\startdata
(1-- 0) R(3) & 1096.73 &   76.6116\\
(1-- 0) P(5) & 1109.31 &  100.7527\\
(1-- 1) R(3) & 1148.70 &  389.9806\\
(1-- 1) P(5) & 1161.82 &  492.7111\\
(1-- 2) R(3) & 1202.45 &  778.3324\\
(1-- 2) P(5) & 1216.07 &  926.0875\\
(1-- 3) R(3) & 1257.83 &  693.4022\\
(1-- 3) P(5) & 1271.93 &  742.2661\\
(1-- 4) R(3) & 1314.62 &  168.5805\\
(1-- 4) P(5) & 1329.14 &  124.0794\\
(1-- 5) R(3) & 1372.49 &   48.7069\\
(1-- 5) P(5) & 1387.37 &  123.9049\\
(1-- 6) R(3) & 1431.01 &  580.3754\\
(1-- 6) P(5) & 1446.12 &  825.4512\\
(1-- 7) R(3) & 1489.57 &  944.7024\\
(1-- 7) P(5) & 1504.76 & 1149.4655\\
(1-- 8) R(3) & 1547.34 &  667.8068\\
(1-- 8) P(5) & 1562.39 &  713.7622\\
(1-- 9) R(3) & 1603.25 &  217.7934\\
(1-- 9) P(5) & 1617.89 &  196.6773\\
(1--10) R(3) & 1655.93 &   23.3209\\
(1--10) P(5) & 1669.77 &   14.8395\\
(1--11) R(3) & 1703.59 &    0.0253\\
(1--11) P(5) & 1716.13 &    0.0397\\
(1--12) R(3) & 1744.02 &    0.0834\\
(1--12) P(5) & 1754.61 &    0.0423\\
(1--13) R(3) & 1774.47 &    0.0090\\
(1--13) P(5) & 1782.18 &    0.0110\\
(1--14) R(3) & 1791.53 &    0.0006\\
\hline
(1-- 0) R(6) & 1109.86 &   69.9280\\
(1-- 0) P(8) & 1130.40 &   88.4209\\
(1-- 1) R(6) & 1161.95 &  377.8252\\
(1-- 1) P(8) & 1183.31 &  447.6342\\
(1-- 2) R(6) & 1215.73 &  809.8802\\
(1-- 2) P(8) & 1237.87 &  873.5052\\
(1-- 3) R(6) & 1271.02 &  797.9877\\
(1-- 3) P(8) & 1293.87 &  729.6058\\
(1-- 4) R(6) & 1327.56 &  247.0670\\
(1-- 4) P(8) & 1351.04 &  133.2556\\
(1-- 5) R(6) & 1385.01 &   22.8098\\
(1-- 5) P(8) & 1408.96 &  111.5518\\
(1-- 6) R(6) & 1442.87 &  554.2507\\
(1-- 6) P(8) & 1467.08 &  800.3663\\
(1-- 7) R(6) & 1500.45 & 1009.0800\\
(1-- 7) P(8) & 1524.65 & 1111.9502\\
(1-- 8) R(6) & 1556.87 &  743.2822\\
(1-- 8) P(8) & 1580.67 &  653.4937\\
(1-- 9) R(6) & 1610.95 &  236.9584\\
(1-- 9) P(8) & 1633.84 &  152.5215\\
(1--10) R(6) & 1661.20 &   20.8714\\
(1--10) P(8) & 1682.49 &    6.1068\\
(1--11) R(6) & 1705.68 &    0.0395\\
(1--11) P(8) & 1724.43 &    0.3495\\
(1--12) R(6) & 1741.95 &    0.0684\\
(1--12) P(8) & 1756.90 &    0.0004\\
(1--13) R(6) & 1766.87 &    0.0179\\
\enddata
\end{deluxetable}

\begin{deluxetable}{ccc}
\tabletypesize{\small}
\tablecaption{H$_{2}$ Model Parameters. \label{table_mod}}
\tablewidth{0pt}
\tablehead{
\colhead{Parameter} &  {NGC 6853}  &  {NGC 3132}}
\startdata
Excitation Source & \ion{H}{1} Ly$\alpha$ & \ion{H}{1} Ly$\alpha$\\
Ly$\alpha$ Doppler shift &  25 km s$^{-1}$ &  30 km s$^{-1}$\\
Ly$\alpha$ FWHM &  0.40 \AA &  0.45 \AA \\
Ly$\alpha$ total intensity &  12944 R  &  217523 R\\
T(H$_{2}$) &  2040 K &  2040 K\\
N(H$_{2}$) &  6.0~$\times$ 10$^{18}$ cm$^{-2}$ &  3.0~$\times$ 10$^{18}$ cm$^{-2}$\\
N(\ion{H}{1}) &  1.0~$\times$ 10$^{14}$ cm$^{-2}$ &  2.0~$\times$ 10$^{16}$ cm$^{-2}$\\
$b$ &  8 km s$^{-1}$ &  9 km s$^{-1}$\\

\enddata
\end{deluxetable}

The model parameters summarized in Table~\ref{table_mod} were adjusted to 
reproduce the 1115~--~1187~\AA\ line strengths observed in the LiF 1B channel. The LiF 2A channel (Figure~\ref{fusespec}) shows a 20\% decrease in the (1--1)~R(3) brightness, which we attribute to calibration offsets. The molecular hydrogen temperature of 2040~K was derived from absorption spectra \citep{mcc06}. The H$_{2}$ absorption is blueshifted with respect to the Ly$\alpha$ profile by about --25~km~s$^{-1}$, value supported by recent studies of NGC~6853 \citep{mcc06}, assuming Ly$\alpha$ at the systemic velocity. The choosen total Ly$\alpha$ brightness constrains the molecular hydrogen column density of 6~$\times$~10$^{18}$~cm$^{-2}$. An H$_{2}$ column density of few~$\times$~10$^{16}$~cm$^{-2}$, as inferred from absorption spectra \citep{mcc06} would require a much higher Ly$\alpha$ brightness.

\subsection{NGC 3132}

In analyzing the NGC~3132 spectrum it is important to account for the background. The 
continuum has a very 
unusual shape, most likely due to the superposition of the
central stars and the inhomogeneous dust distribution within the
nebula.  A detailed analysis of the continuum is beyond the scope 
of this work, but a rough baseline is needed for the
model H$_{2}$ spectrum. We set this background by manipulating 
the original HUT spectrum to extract a smooth continuum curve.
A broad (width~$\sim$~75~\AA) median filter was applied to the data at wavelengths
between 1100 and 1750~\AA\ to remove most of the nebular and
geocoronal emission lines. A boxcar smooth (width~$\sim$~20~\AA) was then 
applied, leaving a background continuum spectrum. Measured brightnesses for the strongest lines are 
given in Table~\ref{table6}. The observed H$_{2}$ emission lines provide a long wavelength confirmation of pumping by Ly$\alpha$ photons.  

\begin{deluxetable}{cccc}
\tabletypesize{\small}
\tablecaption{NGC 3132. \label{table6}}
\tablewidth{0pt}
\tablehead{
\colhead{Line ID} & \colhead{$\lambda_{obs}$ }   &
\colhead{FWHM} & \colhead{Brightness }\\ 
	&	(\AA)		&	(\AA)	&
	(10$^{-6}$ ergs cm$^{-2}$ s$^{-1}$ sr$^{-1}$)	 
}
\startdata
H$_{2}$(1--3)R(3) & 1259.19 &    4.14 &    8.62 $\pm$    0.48\\
H$_{2}$(1--3)P(5) & 1273.38 &    3.57 &   16.81 $\pm$    0.52\\
H$_{2}$(1--6)R(3) & 1431.73 &    2.41 &   11.86 $\pm$    0.53\\
H$_{2}$(1--6)P(5) & 1446.67 &    4.01 &   26.75 $\pm$    0.62\\
H$_{2}$(1--7)R(3) & 1489.85 &    2.27 &   23.12 $\pm$    0.64\\
H$_{2}$(1--7)P(5) & 1505.84 &    2.41 &   22.59 $\pm$    0.58\\
H$_{2}$(1--8)R(3) & 1548.16 &    2.89 &   15.54 $\pm$    0.64\\
H$_{2}$(1--8)P(5) & 1563.10 &    1.92 &   14.35 $\pm$    0.64\\
H$_{2}$(1--9)P(5) & 1617.69 &    1.28 &    7.04 $\pm$    0.75\\
\enddata
\end{deluxetable}

\begin{figure*}
\begin{center}
\epsscale{0.5}
\rotatebox{90}{
\plotone{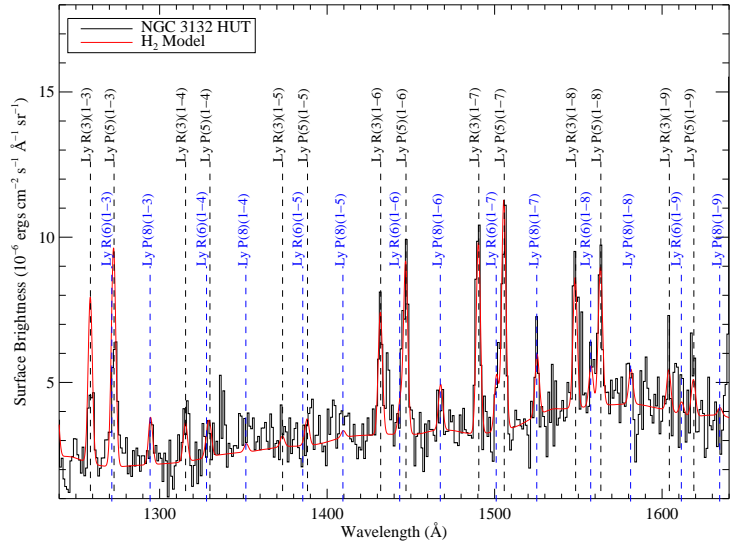} }
\caption{~~HUT spectrum of NGC~3132. The H$_{2}$ model added to an empirical fit to the continuum 
is shown in red (see text). \label{hutspec}  }
\end{center}
\end{figure*}

We use the NGC~6853 model as a starting point for NGC~3132 as a detailed study of the molecular structure of NGC~3132 was not available. Previous observations and models of NGC~3132 \citep{monteiro00,bassgen90} estimate the nebular H$\alpha$ luminosity at $\sim$~few~$\times$~10$^{44}$~photons~s$^{-1}$, too low to predict the observed level of H$_{2}$ fluorescence. This value was derived using a distance of 670~pc, which might be an underestimate. The estimated total H$\alpha$ luminosity of NGC~6853, approximated as an elliptical nebula with semimajor axes of 2\farcm5\ and 4\arcmin, at a distance of 417~pc, is 8.55~$\times$~10$^{46}$~photons~s$^{-1}$. Using for the NGC~3132 0\farcm45$\times$0\farcm7 nebula a total H$\alpha$ luminosity of $\sim$~1~$\times$~10$^{47}$~photons~s$^{-1}$, closer to the value for NGC~6853, we obtain a reasonable fit with 
column densities of 3~$\times$~10$^{18}$~cm$^{-2}$ for the molecular component and 
2~$\times$~10$^{16}$~cm$^{-2}$ for the atomic hydrogen. The molecular hydrogen absorption is blueshifted with respect to the nebular Ly$\alpha$ by --30~km~s$^{-1}$, in agreement with other measurements \citep{monteiro00}. A summary of the parameters used is given in Table~\ref{table_mod}. The overprediction of line
strengths in the 1200~$\leq~\lambda~\leq$~1300~\AA\ region (Figure~\ref{hutspec}) and shortward of 1200~\AA\ (not shown) is caused by an incomplete removal of the
combination of geocoronal Ly$\alpha$ and the (1~--~3) H$_{2}$
emission band together with opacity effects. These shorter wavelength lines connect to more populated lower vibrational levels (Table~\ref{table_branch}), leading to self-absorption \citep{herczeg04,wood02}.

\section{DISCUSSION}

Although H$_{2}$ absorption was observed in both nebulae \citep{mcc06,sterling02}, continuum pumped molecular hydrogen emission was not detected. Predicted continuum pumped UV fluorescence for an H$_{2}$ column density of 7.9~$\times$~10$^{16}$~cm$^{-2}$, measured from absorption spectra \citep{mcc06}, is estimated at a level of $\sim$0.7~$\times$~10$^{-6}$~ergs~cm$^{-2}$~s$^{-1}$~sr$^{-1}$~\AA$^{-1}$, consistent with the noise level in Figure~\ref{fusespec}. However, the larger column density used in the Ly$\alpha$ pumping model, would predict a much higher continuum fluorescence, which is not observed. One possible explanation comes from the dominance of Lyman continuum photons over the 912~--~1110~\AA\ photons which contribute to continuum H$_{2}$ pumping. In the central \ion{H}{2} region of a PN, Lyman continuum photons are readily converted into Ly$\alpha$ line photons, following a series of ionizations, recombinations and cascades. Integrating a stellar model with T$_{eff}$~=~110,000~K, log($g$)~=~6.7, and M$_{\star}$~=~0.56~M$_{\odot}$ \citep{rauch03}, we find that there
are over 16 times more photons~cm$^{-2}$~s$^{-1}$ emitted in the
Lyman continuum (5~--~911.7~\AA) than in the 912~--~1110~\AA\ region. Additionally, the Lyman continuum photons are concentrated into an emission line (FWHM$_{Ly\alpha}$~=~0.4~\AA), compared with the far-UV stellar continuum that spans roughly 200~\AA. For an \ion{H}{2} region with an electron
temperature of 12,000 K (i.e.- NGC 6853; ~\citet{pottasch82}) the recombination efficiency for generating Ly$\alpha$ photons is
roughly 67\% \citep{spitzer78}, which gives us 5.5~$\times$~10$^{45}$~photons~s$^{-1}$. The total Ly$\alpha$ luminosity obtained this way is still low compared to the value of 8.55~$\times$~10$^{46}$ photons~s$^{-1}$ derived from the H$\alpha$ brightness. This shows that the ratio of Ly$\alpha$
photons to 912~--~1110~\AA\ continuum photons might be even higher than estimated.

 However, the dominance of Ly$\alpha$
photons over the 912~--~1110~\AA\ stellar continuum alone is not sufficient to compensate for the larger ground state populations and higher oscillator strengths contributing to other Lyman and Werner transitions out of v\arcsec~=~0. Assuming that molecules reside in high density globules, the covering factor involved in continuum excitation is much smaller than for the diffuse radiation field of scattered Ly$\alpha$. If we let B$_{0}$ be the surface brightness of the exciting field, the total flux absorbed by an H$_{2}$ clump will be B$_{0}\Omega$, where $\Omega$ is the solid angle subtended by the absorber, as seen from the source. The redistributed brightness is then radiated into 4$\pi$, so that we can define an effective surface brightness seen by the absorber as (B$_{0}\Omega$)~/~4$\pi$, where $\Omega$~/~4$\pi$ is the covering factor. For a 10\arcsec~$\times$~10\arcsec\ globule at a 50\arcsec\ separation from the central star, the continuum covering factor is about 0.0032, while for the nebular Ly$\alpha$ it is likely to be unity. This estimate takes into account that, while the continuum photons are coming mainly from the star, the Ly$\alpha$ photons are produced and scattered in the nebula, so that the globules are effectively embedded in a diffuse Ly$\alpha$ radiation field covering all 4$\pi$ sr. 

The broad (98.6~km~s$^{-1}$ FWHM) Ly$\alpha$ line required to pump the H$_{2}$ lines in NGC 6853 is likely to originate in the complex shell structure of the nebula. As shown by \citet{meaburn05}, 
in the case of H$\alpha$ the wide profile is a result of combining the motions of the outer shell 
at 35~km~s$^{-1}$, the inner shell at 13~km~s$^{-1}$ and the bulk motion less than 7~km~s$^{-1}$ in 
the central \ion{He}{2} region. We infer a similarly broad velocity structure should be present in 
NGC~3132 to produce a Ly$\alpha$ of comparable width. A narrower velocity distribution would be allowed if we take into account the redistribution of the Ly$\alpha$ photons in the line wings for large optical depths.

IR observations of NGC 6853 and NGC 3132 made to date \citep{storey84,zuckerman88} conclude that the H$_{2}$ spectrum is shock excited based on the ratios of the S(1) (1--0) and S(1) (2--1) lines. Moreover, this conclusion is supported by the non-detection of continuum pumped UV fluorescence, both in these spectra and in rocket observations of the Dumbbell \citep{mcc01}. However, recent studies \citep{hora99,takami00} show that in dense enough environments the ratio of the S(1) (1--0) and S(1) (2--1) lines can appear thermal, even if the IR emission is excited via UV pumping. In these cases, the study of transitions from higher energy levels becomes important in order to distinguish between the two mechanisms. The detection of H$_{2}$ continuum absorption towards the central star of NGC 6853 shows that continuum pumped fluorescence takes place, although at a level allowed by a non-detection in the $FUSE$ spectra. This by itself does not rule out a partial UV continuum pumping of the IR lines. However, the thermal processes are thought to be dominant, since absorption out of vibrational levels v\arcsec~$>$~2 is not observed, in contrast with reflection nebulae where fourescence is important \citep{meyer01}. The possibility of shock excitation in NGC 6853 and NGC 3132 is also revealed here by the high temperature required by the presence of Ly$\alpha$ pumping. More in-depth analysis of PNe \citep{hora99,davis03} reveal that in most cases where both the rotational and vibrational temperature of H$_{2}$ are around 2000~K, shock heating is the dominant excitation mechanism of IR lines. In order to confirm the shock scenario, we need a better understanding of the gas motions in the two objects, in addition to observations shortward of 2~$\mu$m. 

The observed presence of both ionized species and molecules in the same aperture, and the small covering factor necessary to explain the non-detection of continuum pumped lines confirms the 
current view of molecules residing in globules surrounded by an ionized medium, the source 
of the exciting Ly$\alpha$ radiation \citep{meaburn93,huggins02,speck03}. The presence of globules helps the survival of hydrogen molecules, shielded from photodissociation by hard stellar radiation. Overall, the H$_{2}$ column density derived for the Ly alpha pumping model is much higher than the one derived from absorption spectra (10$^{18}$~cm$^{-2}$ vs 10$^{16}$~cm$^{-2}$), suggesting different properties of the regions they are probing. Taking into account that the globule size in NGC 6853 is likely to be less than 10'', the $FUSE$ observations average between high and low density areas, where the low density regions are thought to have a negligible extinction. The low slit filling fraction for the Ly$\alpha$ pumped radiation in both $FUSE$ and $HUT$ observations allows us to reconcile the derived value of few$\times$10$^{18}$~cm$^{-2}$ for the H$_{2}$ column density with coulmns of 10$^{21}$~cm$^{-2}$ suggested for dense globules from visual exctinction arguments \citep{bohlin78,meaburn93}.  

\subsection{Ro-vibrational Cascade Modified by Ly$\alpha$ Pumping}

\begin{figure*}
\begin{center}
\epsscale{0.65}
\rotatebox{90}{
\plotone{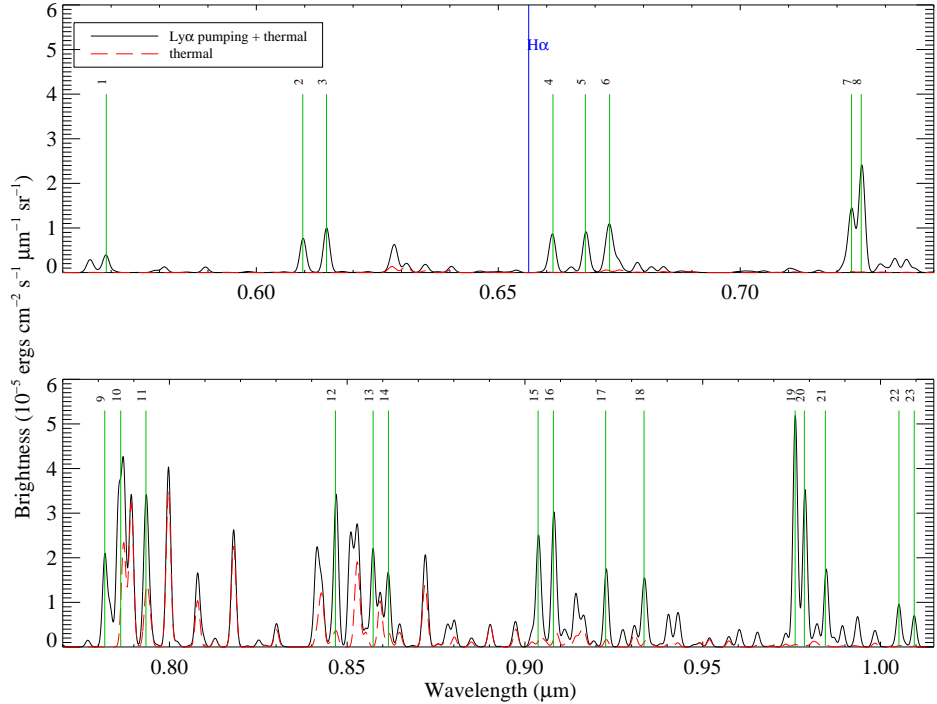} }
\caption{~~Predicted ro-vibrational cascade following Ly$\alpha$ pumping in NGC~6853 (see text). The red dashed line is the ro-vibrational cascade model for a thermal population at 2040~K. The black continuous line has both contributions added. The main contributors to the numbered lines are listed in Table~\ref{table7}. \label{irspec}  }
\end{center}
\end{figure*}

\begin{figure}
\begin{center}
\epsscale{0.45}
\rotatebox{90}{
\plotone{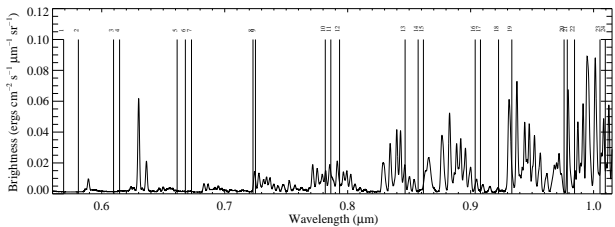} }
\caption{~~Airglow spectrum reconstructed from UVES data \citep{hanuschik03}, with a resolution of 15~\AA. The positions of the strongest lines following Ly$\alpha$ pumping are indicated by the green lines, numbered as in Figure~\ref{irspec}.  \label{UVESspechr}  }
\end{center}
\end{figure}

\begin{figure*}
\begin{center}
\epsscale{0.4}
\rotatebox{90}{
\plotone{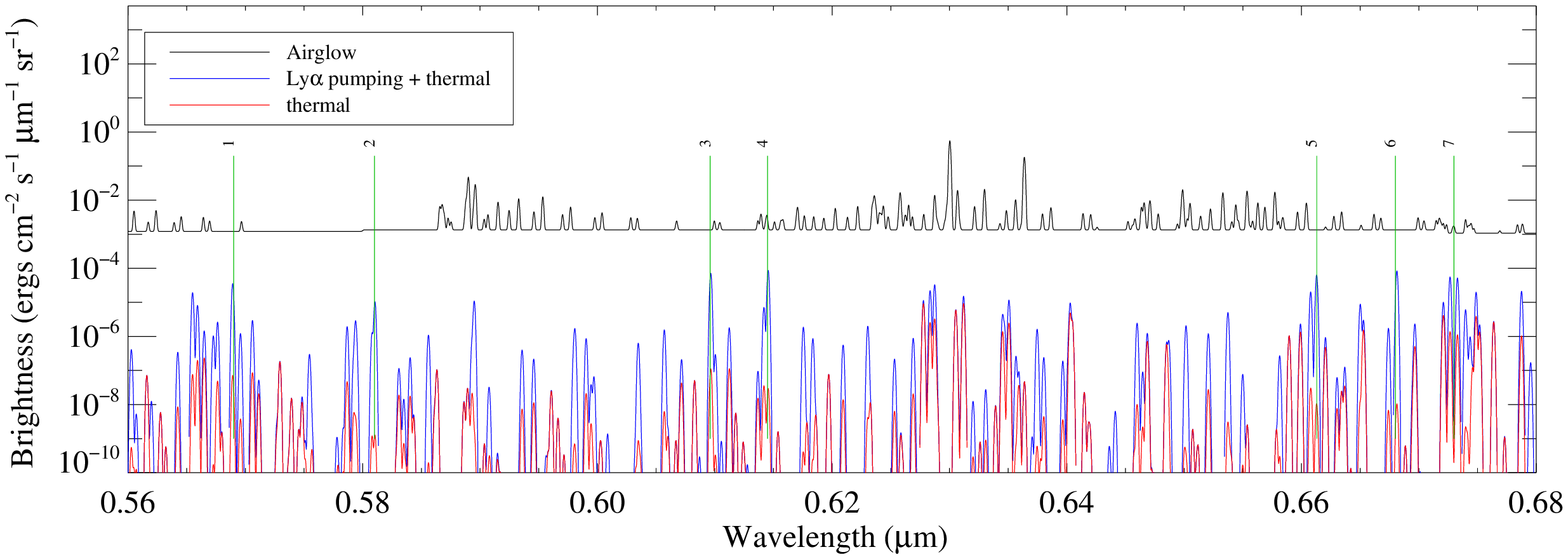} }
\caption{~~Airglow spectrum and ro-vibrational cascade at a resolution of 1.6~\AA, on a logarithmic scale. The black line is the airglow spectrum, while the red and blue are the thermal and co-added (thermal + fluorescent) models, respectively.  \label{UVESspec}  }
\end{center}
\end{figure*}

We constructed the ro-vibrational spectrum as a combination of a non-equilibrium cascade following Ly$\alpha$ pumping and a steady state emission due to the equilibrium distribution of molecules on the $X$$^{1}\Sigma^{+}_{g}$ state levels. For the non-equilibrium cascade, we start with the rates at which the $X$$^{1}\Sigma^{+}_{g}$ levels are populated following Ly$\alpha$ pumping, and evaluate the line strengths according to the branching ratios. Collisional effects are ignored for the purpose of this estimate. As a proxy for the equilibrium populations for the steady state case we use the ground state populations measured from absorption spectra and extrapolated to higher vibrational and rotational levels using a 2040~K thermal distribution with a total H$_{2}$ column density of 7.9~$\times$~10$^{16}$~cm$^{-2}$. For this reason hereafter we will refer to this steady state model as thermal. The transition probabilities from ~\citet{wolniewicz98} are used to derive the output spectrum, shown in Figure~\ref{irspec}. The solid black line contains contributions from both thermal and fluorescent pumping, convolved with a 15~\AA\ Gaussian. Overplotted in red is the thermal contribution alone. The lines showing the largest contribution from fluorescent pumping are numbered, and their principal components are listed in Table~\ref{table7}. The line strengths are given as upper limits and will be affected by collisional de-excitation. Collisional excitation is not important for these high lying vibrational states. Using for the thermal model the same total H$_{2}$ column density of 6~$\times$~10$^{18}$~cm$^{-2}$ used for the Ly$\alpha$ pumping, the lines with the highest contribution from fluorescent cascade will be 1,2,3,4,8,9 and 10, as numbered in Figure~\ref{irspec}.

The ro-vibrational cascade shows most of the Ly$\alpha$ pumping specific features in the visible and near-IR part of the spectrum. We find little deviation from a pure thermal emission longward of $\sim$~1~$\mu$m. As a consequence, line pumped UV fluorescence might be present even when the measured IR line ratios around 2~$\mu$m are consistent with a thermal distribution of the H$_{2}$ molecules in the ground electronic state.

The detection of these lines represents an observational challenge from the ground. In Figure~\ref{UVESspec} the positions of the strongest predicted lines are indicated on the airglow emission spectrum reconstructed from UVES observations \citep{hanuschik03} at a resolution of 15~\AA\ to match the resolution of our model. The lack of airglow features in the 0.577~--~0.583~$\mu$m interval is due to the chip gap in the UVES spectrum. However, \citet{Osterbrock96} shows that there are no lines identified in this region. The ro-vibrational cascade at a level of few~$\times$~10$^{-5}$~ergs~cm$^{-2}$~s$^{-1}$~sr$^{-1}$ $\mu$m$^{-1}$ is about 1000 times smaller than the typical airglow lines and contiuum in the 0.56~--~1.015~$\mu$m interval. A much lower background level and a higher spectral resolution ($\sim$~few~\AA) may allow the detection of the (9--3)~S(4) line at 5810.36~\AA\ and the (8--3)~S(6) line at 6681.40~\AA\ (Figure~~\ref{UVESspechr}).

\begin{deluxetable}{cccc}
\tabletypesize{\small}
\tablecaption{Ro-vibrational cascade, thermal contribution not included. \label{table7}}
\tablewidth{0pt}
\tablehead{
\colhead{Ref No} & \colhead{Line ID}   & \colhead{Wavelength (\AA)}   &
\colhead{Brightness (R)}\\ 
}
\startdata
1  &  (6--1) S(6) &   5689.17 &   0.022\\
2  &  (9--3) S(4) &   5810.36 &   0.007\\
3  &  (7--2) S(4) &   6096.58 &   0.034\\
3  &  (7--2) S(3) &   6096.74 &   0.013\\
4  &  (7--2) S(6) &   6145.49 &   0.058\\
5  &  (8--3) S(3) &   6607.73 &   0.015\\
5  &  (8--3) S(4) &   6612.89 &   0.047\\
6  &  (8--3) S(6) &   6681.40 &   0.061\\
7  &  (5--1) S(4) &   6726.46 &   0.040\\
7  &  (5--1) S(6) &   6732.89 &   0.037\\
8  &  (9--4) S(4) &   7224.75 &   0.024\\
8  &  (6--2) S(4) &   7231.36 &   0.082\\
9  &  (6--2) S(3) &   7250.86 &   0.036\\
9  &  (6--2) S(6) &   7251.65 &   0.134\\
10 &  (7--3) S(4) &   7818.21 &   0.160\\
10 &  (7--3) S(3) &   7833.48 &   0.064\\
11 &  (7--3) S(6) &   7857.03 &   0.240\\
11 &  (3--0) S(6) &   7871.01 &   0.120\\
12 &  (3--0) S(4) &   7934.81 &   0.150\\
13 &  (4--1) S(4) &   8469.30 &   0.255\\
14 &  (8--4) S(6) &   8572.68 &   0.189\\
15 &  (4--1) S(2) &   8615.89 &   0.093\\
16 &  (5--2) S(6) &   9037.36 &   0.203\\
17 &  (5--2) S(4) &   9081.00 &   0.252\\
18 &  (4--1) Q(4) &   9228.52 &   0.034\\
18 &  (5--2) S(2) &   9228.82 &   0.113\\
19 &  (9--5) S(4) &   9333.02 &   0.069\\
19 &  (9--5) S(3) &   9333.40 &   0.021\\
19 &  (5--2) S(1) &   9339.80 &   0.065\\
20 &  (6--3) S(5) &   9758.45 &   0.019\\
20 &  (6--3) S(6) &   9760.54 &   0.486\\
21 &  (6--3) S(4) &   9787.93 &   0.345\\
22 &  (6--3) S(3) &   9847.38 &   0.170\\
23 &  (6--3) S(1) &  10052.00 &   0.092\\
24 &  (5--2) Q(6) &  10095.00 &   0.070\\
\enddata
\end{deluxetable}

Ly$\alpha$ pumping redistributes molecules from the (v\arcsec,J\arcsec) levels (2,6) and (2,5) among the ro-vibrational levels of the ground state. However, at the level of the observed UV fluorescence, a significant deviation from the thermal populations of the (v\arcsec,J\arcsec) levels (2,6) and (2,5) is not found. Correcting for the ro-vibrational transitions that repopulate these levels, we find that the population decrease relative to the thermal distribution ($\Delta$~N$_{vJ}$/N$_{vJ}^{thermal}$) is about 0.0107 and 0.0012 for the (2,6) and (2,5) levels respectively. We find more significant deviations from a thermal population among the rotational levels of the v\arcsec=0 state. These are likely to be affected also by collisional redistribution and do not match the deviations measured from UV absorption spectra \citep{mcc06}. Deviations in the populations of higher vibrational levels (v\arcsec~$>$~2) are predicted to result in a specific signature in the ro-vibrational spectrum (Figure~\ref{irspec}). 

\section{CONCLUSION}

Line pumping by Ly$\alpha$ is required to qualitatively explain the H$_{2}$ emission features in the far UV spectra of  NGC~6853 and NCG~3132. The observed H$_{2}$ fluorescence gives us valuable information about the conditions of the radiation field and dynamics of the molecular gas in planetary nebulae. The input parameters are similar for both objects, suggesting that we do not see an isolated phenomenon. The UV H$_{2}$ spectrum is likely due to thermally excited molecular hydrogen, shielded from the UV continuum inside globules surrounded by strong nebular Ly$\alpha$ emission. We emphasize that while molecular hydrogen pumping by Ly$\alpha$ shows strong features in UV spectra, it could remain unobservable in the IR and visible. UV observations are thus a requirement for investigating this process in various environments where hot H$_{2}$ is exposed to the presence of Ly$\alpha$.  

\acknowledgements

We wish to acknowledge Bill Blair and Paul Feldman for helpful discussion
about UV fluorescence in these objects. KF thanks Aki Roberge for enjoyable
discussion about line pumped fluorescence. $FUSE$ data were
obtained under the Guest Investigator Program (NASA grant NNGO4GK82G)
by the NASA-CNES-CSA $FUSE$ mission, operated by the Johns Hopkins
University. The HUT spectrum of NGC~3132 was downloaded from the MultiMission
Archive at the Space Telescope Science Institute.

\bibliography{ms}

\end{document}